\documentclass[12pt,preprint]{aastex}

\slugcomment{UTTG--06--02}
\begin{document}

\title{ Cosmological Fluctuations Of Small Wavelength }

\author{Steven Weinberg}
\affil{Theory Group, Department of Physics, University of
Texas, 
Austin, TX, 78712}

\email{ weinberg@physics.utexas.edu}

\begin{abstract}
This paper presents a completely analytic treatment of cosmological 
fluctuations whose wavelength is small enough to come within the 
horizon well before the energy densities of matter and radiation 
become equal.  This analysis yields a simple formula for the conventional 
transfer function $T(k)$ at large wave number $k$, which agrees very 
well with computer calculations 
of $T(k)$. It also yields an explicit formula for the microwave background 
multipole 
coefficient $C_\ell$ at very large $\ell$. 
\end{abstract}


\section{Introduction}

The transfer function  gives the wave length dependence of the growth of
perturbations in the cold dark matter density from early times to near
the present.  As such, it plays a central role in theoretical studies
of cosmological structure formation, and it also enters in
the calculation of the microwave background anisotropies of large multipole
number.  For general wave length the transfer function can only be calculated 
numerically.  This paper will present a purely analytic 
solution of the equations
governing the evolution of perturbations in the early universe in the case of 
small wave length,\footnote{In 
speaking of small wavelengths, it is  nevertheless assumed that the 
wavelength is large enough so that the fluctuations are far outside 
the horizon during the era of electron--positron annihilation, and 
large enough so that viscosity and heat conduction are negligible 
until close to the time of recombination, as is the case for all 
fluctuations of physical interest.} which yields a simple formula for
the transfer function in this case, including the numerical parameters
appearing in this formula. 

The most closely related 
previous work seems to be that of Hu \& Sugiyama (1996).  In contrast with their 
work, 
the present paper provides the justification for a crucial step in 
calculating fluctuations in the dark matter density (see footnote 4 
below); it is entirely analytic, even in following perturbations through the 
era of horizon crossing and in analyzing the case of infinite wavelength (which 
is needed
to normalize the transfer function); and  explicit formulas are given 
for the numerical parameters in the 
transfer function at small wavelength and for the cosmic microwave background 
multipole coefficient $C_\ell$ at large $\ell$.

\section{Generalities}

We consider the contents of the universe to consist of radiation plus 
cold dark matter plus baryons (electrons and nuclei). We  include  
neutrinos in the radiation, neglecting the anisotropic part of their 
energy-momentum tensor, which makes possible a purely analytic treatment.
As usual, the cold dark matter is taken to have zero pressure and only 
gravitational interactions.  
For simplicity at first we will 
assume local thermal equilibrium, so that the fractional changes in the baryon 
and
radiation densities are related by 
$\delta\rho_B/\rho_B=3\delta\rho_R/4\rho_R\equiv \delta_R$,
which is a good approximation until late in the matter-dominated era, and we 
will ignore 
the effects of curvature and a cosmological constant, which are 
negligible until near the present.  Later these effects and departures from 
equilibrium will 
be taken into account where they are relevant. 

The evolution of compressional cosmological perturbations under these 
assumptions are governed by the equations:\footnote{These equations 
are a simple extension of  Eqs.~(15.10.50), (15.10.51), and 
(15.10.53) of Weinberg (1972) to the multi-fluid system considered here.}
\begin{eqnarray}
&&\frac{d}{dt}\,\left(a^2\psi\right)=-4\,\pi G 
\,a^2\,\left[\rho_D\delta_D+\left(\frac{8}{3}\rho_R+\rho_B\right)\delta_R\right]
\;,\\
&&\dot{\delta}_D=-\psi\;,~~~~~~~~~\dot{\delta}_R=-\psi+q^2U_R\;,\\
&&\frac{d}{dt}\left[a^5\,\left(\frac{4}{3}\rho_R+\rho_B\right)U_R\right]=-
\frac{4}{9}
a^3\rho_R\delta_R\;.
\end{eqnarray}

Here $q$ is the co-moving wave number; $a(t)$ is the Robertson--Walker 
scale factor; 
$U_R(t)$ is the radiation velocity potential; $\delta_D$  is the fractional 
change $\delta\rho_D/\rho_D$ in the dark matter density $\rho_D$; and dots 
indicate 
ordinary time derivatives.  We are using a synchronous gauge, with 
vanishing  time-time and time-space components of the metric 
perturbation $\delta g_{\mu\nu}$, and with the remaining gauge freedom 
removed by requiring that the cold dark matter velocity vanishes.  In 
this gauge, all effects of gravitational perturbations for compressional modes 
are contained in the field 
$\psi(t)\equiv d( \delta g_{kk}(t)/2a^2(t))/dt$. 

For general wave numbers these equations are too complicated to be 
solved analytically.  However, for large $q$ we can divide the 
evolution of the fluctuations into two {\em overlapping} eras, in each of 
which there are approximations available that allow an analytic 
solution.

\section{Radiation Dominated Era}

First, consider an era so early that $\rho_D$ and $\rho_B$ are much less than 
$\rho_R$, though 
the wavelength may  be inside or 
outside the horizon.  
Here $a\propto \sqrt{t}$ and $t^2=3/32\pi G \rho_R$, and by 
eliminating $U_R$ we obtain a pair of coupled equations for $\delta_R$ 
and $\psi$:
\begin{equation} 
\frac{d}{dt}\left(t\,\psi\right)=-\frac{1}{t}\delta_R\;,~~~~~
\frac{d}{dt}\left(\sqrt{t}\frac{d\delta_R}{dt}\right)+\frac{q^2\sqrt{t
}}{3a^2}\delta_R=-\frac{d}{dt}\left(\sqrt{t}\psi\right)\;.
\end{equation} 
The linear combination of the three independent solutions that grows 
most rapidly for small time is\footnote{Aside from normalization, this solution 
is equivalent to that given for the 
Newtonian potential in a different gauge in Eq.~(48) of Bashinsky 
\& Bertschinger (2002). }
\begin{eqnarray}
&& \delta_R=\frac{2N}{C^2}\left(\frac{2}{\theta}\sin\theta-\left(1-
\frac{2}{\theta^2}\right)\cos\theta-\frac{2}{\theta^2}\right)\;,\\
&&
\psi=N\left(\frac{4}{\theta^3}\sin\theta+\frac{4}{\theta^4}\cos\theta-
\frac{4}{\theta^4}-\frac{2}{\theta^2}\right)\;
\end{eqnarray}
where $N$ is an unknown function of ${\bf q}$  that is presumably fixed during 
the 
era of inflation;
 $\theta\equiv C\sqrt{t}$; and $C$ is the constant $C\equiv [
2q\sqrt{t}/\sqrt{3}a]_{t\rightarrow 0}$.
Also, Eq.(2) gives
\begin{equation}
\delta_D=-
\frac{2N}{C^2}\int_0^{C\sqrt{t}}\left(\frac{4}{\theta^3}\sin\theta+
\frac{4}{\theta^4}\cos\theta-\frac{4}{\theta^4}-
\frac{2}{\theta^2}\right)\,\theta\,d\theta\;.
\end{equation} 
Note that the fractional perturbations $\delta_D$ and $\delta_R$ are 
both of order $\psi/q^2$,
justifying the neglect of the matter term in Eq.~(1) when 
$\rho_R\gg\rho_D$.

For convenience later, it is useful to normalize the Robertson--Walker scale 
factor so that $a=1$ at the time $t_{\rm EQ }$ when the matter density 
$\rho_M\equiv\rho_D+\rho_B$ and  the radiation density $\rho_R$ have a common 
value 
$\rho_{\rm EQ }$.  
Then at early times we have $a\rightarrow (32\pi G\rho_{\rm 
EQ} /3)^{1/4}\sqrt{t}$, and so $C=(q/\sqrt{3})(2\pi G\rho_{\rm EQ }/3)^{-1/4}$.  
Also, $q$ is now defined as the physical wave number $q/a$ at $t=t_{\rm EQ }$.

\section{Deep Inside the Horizon}

Following this is an era in which the dark matter density may not be 
negligible, but the wavelength is well within the horizon.  With the 
physical wave number $q/a$ much greater than the expansion rate, there 
are two kinds of normal mode, that can be calculated using two 
different methods of approximation.  

The first are the ``fast'' modes, for which $d/dt$ acting on 
perturbations gives factors of order $q/a$.   Inspection of 
Eqs.~(1)--(3) shows that there is a solution 
with $\delta_R=O(q\psi)$, $U_R=O(\psi)$, and 
$\delta_D=O(\psi/q)$, so that we can neglect the term $\psi$ on the 
right-hand side of Eq.~(2), and even for $\rho_D>\rho_R$  we can 
neglect the dark matter term on the right-hand side of Eq.~(1).  
Eliminating $U_R$ then gives an equation for $\delta_R$ alone:
\begin{equation} 
\frac{d}{dt}\left((1+R)a\frac{d\delta_R }{dt} 
\right)+\frac{q^2}{3a}\delta_R=0\;,
\end{equation} 
where $R\equiv 3\rho_B/4\rho_\gamma$.  This has the well-known WKB solutions 
(Peebles \& Yu, 1970)
\begin{equation}
\delta_R^{\pm}= (1+R)^{-1/4}\exp\left(\pm 
i\int\frac{q\,dt}{\sqrt{3(1+R)}a}\right)\;,
\end{equation} 
which would be exact for vanishing $R$.

Then there are ``slow'' modes, for which $d/dt$ acting on 
perturbations gives factors of order $1/t$.
Inspection of Eqs.~(2)--(3) shows that in this case 
there is a solution with $\delta_D=O(\psi)$, $\psi \simeq q^2U_R$. and  
$\delta_R=O(U_R)=O(\psi/q^2)$.   It follows that even for 
$\rho_D<\rho_R$  we can neglect the radiation term on the right-hand 
side of Eq.~(1), so that after eliminating the field $\psi$ we 
have
\begin{equation} 
 \frac{d}{dt}\,\left(a^2\frac{d\delta_D}{dt}\right)=4\,\pi G 
\,a^2\,\rho_D\delta_D\;.
\end{equation} 
It is convenient to convert the independent variable from $t$ to $a$, using the 
Friedmann equation
\begin{equation} 
\frac{\dot{a}^2}{a^2}=\frac{8\pi G}{3}\left(\rho_M +\rho_R\right)= 
\frac{8\pi G\rho_{\rm EQ}}{3}\left(a^{-3}+a^{-4}\right)\;
\end{equation} 
so that Eq.~(10) reads\footnote{ Hu and Sugiyama (1996) pointed out that this 
equation leads to a 
transfer function with the asymptotic form $\ln k/k^2$, but it has not 
been clear why it is  legitimate in deriving Eq.~(12) to neglect fluctuations in 
the radiation energy density as a contribution to the source of the 
gravitational field 
during the radiation-dominated era.  Eq.~(12) was first derived by  
M\'{e}sz\'{a}ros (1974), who simply ignored fluctuations
in the radiation density.   Groth and Peebles (1975) neglected fluctuations 
in the radiation density on the
grounds that the wavelength is much less than the Jeans length, which  for 
radiation 
is the horizon, but the relevance
of the Jeans length in an expanding universe containing both radiation and dark 
matter 
is not clear.  In their Appendix B, Hu and Sugiyama (1996)
neglected the contributions of perturbations in the dark matter density to the 
gravitational
field at early times, and showed that then the contribution of perturbations
in the radiation density to the gravitational field are also negligible.  But 
this does not 
justify the use of equation (12).  For this, it is necessary to show that 
perturbations
in the radiation density make negligible contributions to the gravitational 
field when the
contributions of the dark matter perturbation are {\em not} negligible, as is 
the case
late in the radiation dominated era and during the cross-over from radiation to 
matter dominance. Liddle and Lyth (2000) on p. 107 attempted to explain the 
neglect 
of perturbations in the radiation energy density in  Eq.~(12) by claiming that 
Silk damping makes 
these perturbations decay away.  This is incorrect.  For wavelengths 
of physical interest Silk damping is negligible during the 
radiation-dominated era and through the time of radiation-matter equality.  
(This has been
acknowledged by Liddle and Lyth in an erratum: 
 star-www.cpes.sussex.ac.uk/~andrewl/infbook/errata.html.)
 The 
neglect of perturbations in the radiation density in Eq.~(12) is 
explained by counting powers of $1/q$ as done here, and it applies 
only to the slow mode part of the solution; in the fast mode it is the 
perturbations in the {\em dark matter} density that become negligible for small 
wavelength.}
\begin{equation}
a(1+a)\frac{d^2\delta_D}{da^2}+\left(1+\frac{3a}{2}\right)\frac{d\delta_D}{da}-
\frac{3}{2}(1-\beta)\,\delta_D=0\;,
\end{equation}
where $\beta\equiv\rho_B/\rho_M=\Omega_B/\Omega_M$.
The 
independent  solutions of Eq.~(12) for $\beta =0$ were given by M\'{e}sz\'{a}ros 
(1974) and  
Groth \& Peebles (1975):
\begin{equation}
f_1=1+\frac{3a}{2}\;,~~~~~~f_2=\left(1+\frac{3a}{2}\right)\,\ln\left(\frac{\sqrt
{1+a}+1}{\sqrt{1+a}-1}\right)-3\sqrt{1+a}\;.
\end{equation}
Hu and Sugiyama (1996) have given the solutions for general $\beta$  in terms of 
hypergeometric functions, but the necessity of matching these solutions to those 
that apply after recombination leads to an extremely complicated formula for the 
transfer function, which obscures the dependence of the result on the baryon 
density.  Here we will assume that $\beta$ is small, though not entirely 
negligible, and work with solutions valid only to first order in $\beta$.  The 
first-order solutions of Eq.~(12) with the same behavior  for $a\ll 1$ as the 
zeroth order solutions $f_1$ and $f_2$ are:
\begin{equation}
\delta_D^{(1,2)}(a)=f_{1,2}(a)-\frac{3\beta}{2}\int_0^a\left[f_1(a)f_2(b)-
f_2(a)f_1(b)\right]\frac{f_{1,2}(b)\,db}{\sqrt{1+b}}\;.
\end{equation}

By applying Eqs.~(2) and (3), we can find the fast mode solutions for $U_R^\pm$ 
and $\delta_D^\pm$ from Eq.~(9)  and the slow mode solutions for $U_R^{(1,2)}$ 
and $\delta_R^{(1,2)}$ from Eq.~(14).  These four modes a complete set of 
solutions of the 
fourth-order system of equations (1)--(3) up to the time of recombination for 
$q/a\gg \dot{a}/a$.  
The physical solution is a linear combination of these four modes, to be found 
by matching their behavior for $a\ll 1$ to that found in Section 3.

\section{Matching}

Fortunately, for small wavelength there is an overlap in the two eras 
in which we have found solutions for $\delta_D$, etc., satisfying {\em both} 
conditions 
$q/a\gg \dot{a}/a$  and $\rho_M\ll\rho_R$.   In this period 
$C\sqrt{t}\gg 1$, and Eq.~(5) gives the oscillating part of the fractional 
perturbation in the radiation density as
$\delta_R=-(2N/C^2)\cos C\sqrt{t}$, which for $\rho_M\ll\rho_R$ fits smoothly 
with the 
linear combination of the fast solutions (9) for $q/a\gg \dot{a}/a$:
\begin{equation}
\delta^{\rm fast}_R=-\frac{2N}{(1+R)^{1/4}C^2}\cos \left(\int_0^t 
\frac{q\,dt}{\sqrt{3(1+R)}a}\right)= -\frac{2N\sqrt{6\pi G\rho_{\rm EQ 
}}}{(1+R)^{1/4}q^2}\cos \left(\int_0^t 
\frac{q\,dt}{\sqrt{3(1+R)}a}\right)\;,
\end{equation} 
from which we also find, to leading order in $1/q$,
\begin{equation}
\psi^{\rm fast}=\frac{3Na(1+
R)^{1/4}(2+R) \sqrt{2\pi G\rho_{\rm EQ }}}{q^3t^2}\,\sin \left(\int_0^t 
\frac{q\,dt}{\sqrt{3(1+R)}a}\right)\;,
\end{equation}
\begin{equation}
\delta^{\rm fast}_D=\frac{3 N a^2 (1+R)^{3/4}(2+R) \sqrt{6\pi G\rho_{\rm EQ 
}}}{q^4t^2}\,\cos \left(\int_0^t 
\frac{q\,dt}{\sqrt{3(1+R)}a}\right)\;.
\end{equation}
and
\begin{equation} 
U_R^{\rm fast}=\frac{2N\sqrt{2\pi G\rho_{\rm EQ }}}{q^4 a(1+R)^{3/4}}\,\sin 
\left(\int_0^t 
\frac{q\,dt}{\sqrt{3(1+R)}a}\right)\;.
\end{equation}

To find the coefficients in the slow modes, we note that the limit of 
Eq.~(7) for $C\sqrt{t}\gg 1$ is
\begin{equation} 
\delta_D\rightarrow \frac{4N}{C^2}\left(-\frac{1}{2}+\gamma+\ln 
C\sqrt{t}\right)=\frac{4N\sqrt{6\pi G\rho_{\rm EQ }}}{q^2}\left(-
\frac{1}{2}+\gamma+\ln 
\frac{aq}{\sqrt{8\pi G\rho_{\rm EQ }}}\right)\;,
\end{equation} 
where $\gamma=0.5772\dots$ is the Euler constant.  
For $a\ll 1$, the 
solutions (14) become 
\begin{equation} 
\delta^{(1)}_D\rightarrow 1\;,~~~~~\delta^{(2)}_D\rightarrow -\ln(a/4)-3\;.
\end{equation} 
The linear combination of these solutions that fits smoothly with 
Eq.~(19) is then
\begin{eqnarray} 
\delta_D^{\rm slow}&=&\frac{4N}{C^2}\left\{\left[-
\frac{7}{2}+\gamma+\ln\left(\frac{2\sqrt{3}C^2}{q}\right)\right]\delta
_D^{(1)}
-\delta_D^{(2)}\right\}\nonumber\\
&=& \frac{4N\sqrt{6\pi G\rho_{\rm EQ}}}{q^2}\left\{\left[-
\frac{7}{2}+\gamma+\ln\left(\frac{2q}{\sqrt{2\pi G\rho_{\rm 
EQ}}}\right)\right]\delta_D^{(1)}
-\delta_D^{(2)}\right\}\;.
\end{eqnarray}
The slow part of the velocity potential and radiation density   are given by 
Eqs.~(2), (3), and (21) as 
\begin{equation} 
U_R^{\rm slow}=\psi^{\rm slow}/q^2=-\dot{\delta}_D^{\rm slow}/q^2
\end{equation} 
\begin{equation} 
\delta_R^{\rm slow}=-3a^2\frac{d}{dt}\left[a(1+R) U^{\rm slow}_R\right]\;.
\end{equation} 
The full solution up to the time of recombination is given by
$\delta_D=\delta_D^{\rm fast}+\delta_D^{\rm slow}$ and likewise for $U_R$ and 
$\delta_R$.

Eq.~(17) shows that the fast part of $\delta_D$ is smaller than the slow part 
(21) by a 
factor of order $1/q^2t^2$, so that for small wavelengths the full perturbation 
to the dark matter density is given by Eq.~(21) from the time that 
$q/a$ becomes much greater  than $\dot{a}/{a}$,  and even after the 
energy densities of matter and radiation become comparable, up to the time of 
recombination.  But this is not 
true of the radiation perturbations. 
Comparison of Eqs.~(22) and (23) with (18) and (15) shows that for large $q$ the 
perturbations to the radiation velocity potential and density are dominated by 
the fast 
mode, by one and two 
factors of $q$, respectively. 

\section{The Transfer Function}

The transfer function $T(k)$ is properly defined as the growth of the 
total matter density perturbation for a given present physical wave 
number $k\equiv q/a(t_0)=q(1+z_{\rm EQ })$, from early in the 
radiation-dominated era 
to late in the matter dominated era, relative to the growth that 
occurs in the same time interval for zero wave number.  We must therefore now 
project the solution we have found for the density perturbations forward into 
the era following the time of recombination.
In this era the baryonic perturbation is no longer suppressed by radiation 
pressure, 
and so it follows the same equation as the dark matter perturbation:
\begin{equation}
a(1+a)\frac{d^2\delta_B}{da^2}+\left(1+\frac{3a}{2}\right)\frac{d\delta_B}{da}=a
(1+a)\frac{d^2\delta_D}{da^2}+\left(1+\frac{3a}{2}\right)\frac{d\delta_D}{da}=
\frac{3}{2}\,\delta_M\;,
\end{equation}
where $\delta_M\equiv \delta\rho_M/\rho_M=(1-\beta)\delta_D+\beta\delta_B$.  
This 
does not mean that $\delta_B$ and $\delta_D$ are equal, for they satisfy 
different initial conditions at recombination.  But from a linear combination of 
these equations for $\delta_D$ and $\delta_B$ we find that 
\begin{equation}
a(1+a)\frac{d^2\delta_M}{da^2}+\left(1+\frac{3a}{2}\right)\frac{d\delta_M}{da}-
\frac{3}{2}\,
\delta_M=0\;,
\end{equation}
This has the solutions (13).  To find the correct linear combination of these 
solutions, we note that $\delta_B$ and $\dot{\delta}_B$ vanish to leading order 
in $1/q$ at recombination, so $\delta_M$ and and its first derivative at 
recombination must  
respectively equal $(1-\beta)\delta_D$ and its first derivative.  The total 
matter 
density perturbation after recombination is the linear combination of the 
solutions (13) that matches in this way with the solution (21):
\begin{equation}
\delta_M(a)= \frac{4N\sqrt{6\pi G\rho_{\rm 
EQ}}}{q^2}\Big(A_1f_1(a)+A_2f_2(a)\Big)\;,
\end{equation}
where 
\begin{eqnarray}
A_1(q)&=&\left[-\frac{7}{2}+\gamma+\ln\left(\frac{2q}{\sqrt{2\pi G \rho_{\rm 
EQ}}}\right)\right]\left(1-\beta-\beta\,{\cal I}_{12}\right)-\beta\,{\cal 
I}_{22}\;,
\\
A_2(q)&=&-1+\beta-\beta\,{\cal I}_{12}-\beta\,{\cal I}_{11}\left[
-\frac{7}{2}+\gamma+\ln\left(\frac{2q}{\sqrt{2\pi G \rho_{\rm 
EQ}}}\right)\right]\;,
\end{eqnarray}
\begin{equation}
{\cal I}_{ij}\equiv 
\frac{3}{2}\int_0^{a_R}\frac{f_i(a)\,f_j(a)\,da}{\sqrt{1+a}}\;.
\end{equation}
Near the present, where $a\gg 1$, the matter density fluctuation goes to
\begin{equation}
\delta_M\rightarrow \frac{18A_1Na}{q^2}\sqrt{\frac{2\pi 
G\rho_{\rm EQ}}{3}}\;.
\end{equation}
We can find the behavior of $\delta_M$ early in the matter-dominated 
era by taking $C\sqrt{t}\ll 1$ in Eqs.~(5) and (7): 
\begin{equation}
\delta_M\rightarrow Nt/2\;.
\end{equation} 

Eqs.~(30) and (31) must be compared with the growth of $\delta_M$ for 
$q=0$.  In this case Eq.~(2) gives
$\delta_R=\delta_D=-\dot{\psi}$, so Eq.~(1) becomes
\begin{equation}
\frac{d}{dt}\left(a^2\frac{d\delta_M}{dt}\right)=4\pi G a^2\rho_{\rm 
EQ}\left(\frac{1}{a^3}+\frac{8}{3a^4}\right)\, \delta_M 
\end{equation} 
The solution of Eqs.~(32) and (11) that has the same behavior for 
$a\rightarrow 0$ as Eq.~(31) is
\begin{equation}
\delta_M=\frac{N}{5a^2}\sqrt{\frac{3}{2\pi G \rho_{\rm 
EQ}}}\left(16+8a-2a^2+a^3-16\sqrt{1+a}\right)
\end{equation} 
This 
has an asymptotic behavior for $a\gg 1$:
\begin{equation}
\delta_M\rightarrow\frac{Na}{5}\sqrt{\frac{3}{2\pi G \rho_{\rm EQ}}}
\end{equation}
The transfer function $T$  then has an asymptotic behavior for 
large wave number given by the ratio of Eqs.~(30) and (34):
\begin{equation}
T\rightarrow \frac{60 \pi G\rho_{\rm EQ }}{ q^2 }A_1(q)\;,
\end{equation} 
with $A_1(q)$ given by Eq.~(27).
At late times the growth of $\delta_M$ may be affected by a cosmological 
constant or spatial curvature, but these effects are independent of wave number, 
and therefore cancel in the transfer function.

For $\rho_B\ll\rho_D$ we can neglect $\beta$,  so Eqs.~(35) and 
(27) give a transfer function
\begin{equation}
 T\rightarrow \frac{60 \pi G\rho_{\rm EQ }}{ q^2}\left[-
\frac{7}{2}+\gamma+\ln\left(\frac{2q}{\sqrt{2\pi G \rho_{\rm EQ 
}}}\right)\right]\;.
\end{equation} 
This can be put in more familiar terms by using the 
relations $\rho_{\rm EQ}=(3H_0^2\Omega_M/8\pi G)(1+z_{\rm EQ})^3$, 
$q=k(1+z_{\rm EQ})$, and
$1+z_{\rm EQ}=\Omega_M/\Omega_R$, which give the transfer function in terms of 
the present wave number $k$:
\begin{equation}
T(k)\rightarrow\frac{45 \Omega_M^2H_0^2}{2\Omega_Rk^2}\left[-
\frac{7}{2}+\gamma+\ln\left(\frac{4k\sqrt{\Omega_R}}{\Omega_MH_0\sqrt{
3}}\right)\right]=\frac{\ln(
2.40\,Q)}{(4.07\, Q)^2}\;,
\end{equation} 
where $Q\equiv k ({\rm Mpc}^{-1})/\Omega_Mh^2$, and in the final 
expression we use $\Omega_Rh^2=4.15\times 10^{-5}$.
This may be compared with the BBKS numerical fit (Bardeen et al. 1986) to 
computer 
calculations of the transfer function: 
\begin{equation} 
T(k)\simeq \frac{\ln(1+2.34Q)}{2.34Q} 
\left[1+3.89Q+(16.1Q)^2+(5.46Q)^3+(6.71Q)^4
\right]^{-1/4}.
\end{equation} 
This goes to $\ln (2.34 Q)/(3.96\,Q)^2$ for large $Q$, in very good 
agreement with Eq.~(37).  Our simple calculation thus accounts not 
only for the form of the transfer function for large wave numbers, but 
also for its numerical parameters.

(Though it is not relevant to the present work, it may be noted that the BBKS 
formula cannot be taken seriously for small values of $Q$, for it
has unphysical terms that are linear in $Q$ at $Q\rightarrow 0$.  Analyticity in 
the three-vector ${\bf k}$ requires that in this limit $T(k)$ should 
be a power series in $k^2$, or equivalently in $Q^2$.)

To assess the effect of a non-zero baryon number, we note that, to first order 
in $\beta=\Omega_B/\Omega_M$,  the general formula (35) may be put in the form
\begin{equation}
T\rightarrow \frac{\ln \Big(2.40\,Q\,(1+\beta{\cal 
I}_{22})\Big)}{\left[4.07\,Q\,\Big(1+\beta(1+{\cal I}_{12})/2\Big)\right]^2}
\end{equation} 
We need values for the integrals ${\cal I}_{ij}$ defined by Eq.~(29).  The upper limit on the integrals (29) is 
$a_R=(1+z_{\rm EQ})/(1+z_R)$.  
The redshift $z_R$ at recombination has only a very weak dependence on 
cosmological parameters, and will be taken here to have the fixed value 
$z_R=1100$.  The redshift $z_{\rm EQ}$ at matter-radiation equality is given by 
$1+z_{\rm EQ}=\Omega_M/\Omega_R$, so taking $\Omega_Rh^2=4.15\times 10^{-5}$, we 
have $a_R=21.9\,\Omega_Mh^2$. The integral ${\cal I}_{12}$ is given by
$$
{\cal I}_{12}(a_R)=\frac{3}{20}\left[-22a_R-18a_R^2+4\ln\left(\frac{a_R}{4}\right)
+\sqrt{1+a_R}\left(4+8a_R+9a_R^2\right)\,
\ln\left(\frac{\sqrt{1+a_R}+1}{\sqrt{1+a_R}-1}\right)\right]
$$
 The integral ${\cal I}_{22}$ is given by a lengthy expression involving Spence functions,  but it converges so rapidly for likely values of $a_R$ that for practical purposes we can use the value  for $a_R$ infinite:
$${\cal I}_{22}(\infty)=2\pi^2/5-3= 0.947842\;.$$
  For instance, for $a_R=4.38$ (corresponding to $\Omega_Mh^2=0.2$), we have ${\cal I}_{22}= 0.9470$.

Eq.~(39) agrees very well for large $k$ with the numerical results of Holtzman 
(1989).  For $\Omega_Mh^2=0.2$ (the most plausible of the values considered by 
Holtzman) Eq.~(39) becomes
$$
T\rightarrow \frac{\ln 
\Big(12.0\,k\,(1+0.947\,\beta)\Big)}{\left[20.35\,k\,\Big(1+1.377\,\beta
\Big)\right]^2}\;,
$$
with $k$ in Mpc$^{-1}$.
Table 1 compares the results given by this formula with the numerical results 
given by Holtzman (1989) for $\beta\equiv \Omega_B/\Omega_M$ equal to 0.01 and 
0.1, and for various values of $k$.  As can be seen, for these parameters the asymptotic formula (39) gives a pretty good approximation to the  
numerically calculated results for $k>0.5\, {\rm Mpc}^{-1}$, and the numerically calculated results converge rapidly to Eq.~(39) for larger values of $k$.  
But Holtzman warns that his result should not be used for $k>3.09$ Mpc$^{-1}$, while the 
results obtained from Eq.~(39) presumably become increasingly more accurate for 
larger values of $k$

We see that for any plausible value of $\Omega_Mh^2$,  the effect of a small 
baryon density on the argument of the logarithm is to replace the parameter 
$\Omega_Mh^2$ in the definition of $Q$ with $\Omega_Mh^2(1-.95\beta)$, while   
for $\Omega_Mh^2$ in the range of 0.12 to 0.2, the effect of a small baryon 
density on the denominator of $T(k)$ is to replace 
the parameter $\Omega_Mh^2$ in the definition of $Q$ with $\Omega_Mh^2(1-
\zeta\beta)$, with $\zeta\equiv (1+{\cal I}_{22})/2$  in the range of 
1.24 to 1.38.   
These results throw some light on a series of attempts to 
correct the transfer function for the effects of baryon density by re-scaling 
the definition of Q (usually called q) in the BBKS formula (38). 
Various authors attempted to correct for the baryon density  by 
replacing $\Omega_Mh^2$
in the denominator of $Q$  with a factor $\Omega_Mh^2\exp(-
2\Omega_B)$
(Peacock \& Dodds 1994), or 
with $\Omega_M h^2 \exp(-\Omega_B-\sqrt{2h}\Omega_B/\Omega_M)$ (Sugiyama 1995) 
or with $\Omega_M h^2\exp(-\Omega_B-
\Omega_B/\Omega_M)$
(Liddle et al. 1996).  (For a more detailed study of the 
effects of a finite 
baryon-to-dark matter ratio on the transfer function, see  Eisenstein \& Hu 
(1998).)    
Of course there is no reason why the baryon density 
should enter only in the definition of $Q$, and Eq.~(39) shows that it does not.  
But even without any detailed calculations, it is evident that these correction 
factors are physically impossible.  The transfer function is defined with no 
reference to the present moment, except that it is conventionally written as a 
function of the present wave number $k$.  It depends on 
$\Omega_Mh^2$ and $\Omega_Rh^2$, which enter in the formulas for $k/q$ 
and $\rho_{\rm EQ}$, and it can (and does) have an additional dependence on the 
constant ratio of the energy densities of baryons and all matter, which is 
equal to $\Omega_B/\Omega_M$, but there is no way that it can depend 
separately on $\Omega_B$ or $\Omega_M$ or $h$.   What we  have found here is 
that for large wave number, the effect of a small baryon density can be crudely 
taken into account by replacing the parameter $\Omega_Mh^2$ in the definition of 
$Q$ with $\Omega_Mh^2(1-\zeta\Omega_B/\Omega_M)\simeq\Omega_Mh^2\exp(-\zeta\Omega_B/\Omega_M)$, with $\zeta$ roughly equal 
to unity for likely values of $\Omega_Mh^2$.

\section{ Microwave Background Anisotropies}

The  fractional temperature fluctuation in a direction $\hat{n}$ 
takes the general form (apart from late-time effects):
\begin{equation}
\frac{\Delta T(\hat{n},z)}{T} 
=\int p(z)\,dz  \int d^3k\,e^{i(1+z)d_A(z)\hat{n}\cdot{\bf 
k}}\,\epsilon_{\bf k}\left[F(k, z)+i(\hat{n}\cdot\hat{k}) G(k, z) \right]\;,
\end{equation}
where $p(z)\,dz$ is the probability that last scattering will occur between 
redshifts $z$ and $z+dz$; $d_A(z)$ is the angular diameter distance to redshift 
$z$; and $\epsilon_{\bf k}$ is a primordial fluctuation amplitude, defined as 
proportional to $N({\bf k})$, with a coefficient to be chosen below. In the 
synchronous gauge and hydrodynamic approximation used here, and now making the 
further approximation that dark matter 
dominates the energy density at last scattering, the form factors $F$ and $G$  
in Eq.~(40) are given by (Weinberg 2001):
\begin{eqnarray}
\epsilon F&=&\frac{1}{3}\phi+\frac{1}{3}\delta_R \;,\\
\epsilon G&=&-aqU_R+ qt\phi/a\;,
\end{eqnarray}
where $\phi=-4\pi G \rho_R\delta_R a^2/q^2$ is the Newtonian potential produced 
by dark matter density fluctuations.  The first and second  terms in  $F$ arise 
from the Sachs--Wolfe effect and intrinsic temperature fluctuations, 
respectively.  The form factor $G$ arises from the Doppler effect, with its 
first and second terms contributed by velocities produced by pressure and 
gravitational forces, respectively.

The conventional multipole coefficient $C_\ell$ is given in general by the 
familiar formula
\begin{equation}
C_\ell=16\pi^2\int_0^\infty {\cal P}(k)\,k^2\,dk\left|\int dz\, 
p(z)\,\left[j_\ell\Big(kr(z)/H_0 \Big)\,F(k,z)
 + j'_\ell\Big(kr(z)/H_0 \Big)\,G(k,z)\right]\right|^2\;,
\end{equation} 
where ${\cal P}(k)$ is the power spectral function, defined by
\begin{equation} 
\langle \epsilon_{\bf k}\epsilon_{\bf k'}\rangle ={\cal P}(k)\delta^3({\bf 
k}+{\bf k}')\;,
\end{equation} 
and 
\begin{equation}
r(z)\equiv (1+z)d_A(z)H_0=\int_{1/(1+z)}^1\frac{dx}{\sqrt{\Omega_\Lambda 
x^4+\Omega_M x}}\;.
\end{equation} 
To fix the normalization of $\epsilon_{\bf k}$, we note that for small values of 
$\ell$ (say, $\ell<10$)
the large value of $d_A$ makes the spherical Bessel functions in Eq.~(43) 
oscillate rapidly except for small values of $k$, so $C_\ell$ is dominated for 
small $\ell$ by the Sachs--Wolfe term in $F(k)$, for which Eqs.~(34) and (41) 
give  the 
$z-independent$ small-$k$ behavior
\begin{equation} 
\epsilon F(k,z)\rightarrow -\frac{4\pi G \rho_R a^3 N({\bf q})}{5q^2\sqrt{6\pi G 
\rho_{\rm EQ }}}\;.
\end{equation} 
We therefore define $\epsilon_{\bf k}$ by
\begin{equation}
\epsilon_{\bf k}=-\frac{4\pi G \rho_R a^3 }{5q^2\sqrt{6\pi G \rho_{\rm EQ }}} 
N({\bf q})\;,
\end{equation} 
so that $F(0,z)=1$.  With this normalization, a Harrison--Zel'dovich power 
spectral function ${\cal P}(k)=B k^{-3}$ gives $C_\ell=8\pi^2 B\ell(\ell+1)$ for 
small $\ell$.

For large $\ell$, the integral over $k$ in Eq.~(43) is dominated by large wave 
numbers.  In this case, the Sachs--Wolfe term in Eq.~(41) receives a 
contribution of order $1/k^4$ from the slow mode part (21) of $\delta_D$ and of 
order $1/k^6$ from the fast mode part (17). The intrinsic fluctuation term in 
Eq.~(41) receives a contribution of order $1/k^2 $ from the fast mode term (15) 
in $\delta_R$,  and of order $ 1/k^4$ from the slow mode term (23).
 The  slow mode parts of the two terms in the Doppler form factor (42) cancel, 
leaving the contribution of the fast mode term (18) in $U_R$, which is of order
$1/k^3$. 
We conclude from this that in the absence of dissipative effects, the 
temperature fluctuation is dominated for large $k$ by the fast-mode part of the  
intrinsic temperature fluctuation.  

But for very large $k$ the rapidly oscillating fast mode is killed by Silk 
damping (i., e., photon viscosity and heat conduction) and Landau damping 
(cancelations due to large changes in the phase of the fast modes over the range 
of redshifts at which last scattering may occur).  As pointed out by Hu and 
Sugiyama(1996), for $\ell$ greater than about 4,000 the dominant contribution to 
$C_\ell$ arises from the non-oscillatory terms in the perturbations.  These 
terms, which are contributed by both the Sachs--Wolfe effect and the intrinsic 
temperature fluctuations, can be taken from Eqs.~(83) and (84) of Weinberg 
(2001), with 
the damped terms neglected and an extra factor $T(k)$ supplied, because here we 
are dealing with wavelengths that come into the horizon during the radiation 
dominated era.  This 
gives 
\begin{equation}
F(k,z) \rightarrow-3R(z)\, T(k)\;,~~~~~~~~~~G(k,z) \rightarrow 0\;,
\end{equation}
so Eq.~(43) becomes
\begin{equation}
C_\ell=144\pi^2\int_0^\infty {\cal P}(k)\, T^2(k) \,k^2\,dk\left|\int dz\, 
p(z)\, R(z)\,j_\ell\Big(kr(z)/H_0\Big) 
\right|^2\;.
\end{equation} 
To do the double integral over $z$ and $k$, we use an approximation of Hu and 
White (1996).  The 
last-scattering probability distribution $p(z)$ is sharply peaked around a mean 
value $z_L\simeq 1,100$, while for sufficiently large $\ell$ the spherical 
Bessel function is even more sharply peaked at a value $\ell+1/2$ of its 
argument.  We therefore set $z$ at a value where $kr(z)/H_0=\ell+1/2$ everywhere 
but in the argument of $j_\ell$, 
and integrate over the argument of $j_\ell$ with $k$ fixed, after which  we  set 
$k=(\ell+1/2)H_0/r(z_L)$ everywhere but in the argument of $p(z)$, and integrate 
over that argument:
\begin{eqnarray}
C_\ell &\rightarrow & 144\pi^2 H_0^3\,{\cal P}\left(\frac{(\ell+1/2)H_0}{r(z_L) 
}\right) \,T^2\left(\frac{(\ell+1/2)H_0}{r(z_L) 
}\right)\,R^2(z_L)\,\frac{\ell+1/2}{r^2(z_L)r'(z_L) }\nonumber\\&&\times                                                                                                                                                                                                                                                                                                                                                                                                      
\int p^2(z)\,dz\,\left|\int_0^\infty  j_\ell(s)\,ds 
\right|^2\nonumber\\
&\rightarrow &
\frac{36\pi^{5/2}H_0^3(1+z_L)^{3/2}\sqrt{\Omega_M}}{r^2(z_L)\sigma}\,{\cal 
P}\left(\frac{\ell\,H_0}{r(z_L)}\right)\,T^2\left(\frac{\ell\,H_0}{r(z_L)}\right
)\,R^2(z_L)\;,
\end{eqnarray}
where $\sigma$ is defined by
\begin{equation}
\int p^2(z)\,dz\equiv \frac{1}{2\sqrt{\pi}\sigma}\;,
\end{equation} 
so that $\sigma$ is the standard deviation if $p(z)$ is Gaussian.  For instance, 
for a straight spectrum with ${\cal P}(k)\propto k^{-2-n_s}$, Eq.~(45) gives 
$C_\ell\propto \ell^{-6-n_s}\ln^2\ell$.  Unfortunately the interposition of 
foreground objects makes it unlikely that this can be measured.

I am grateful for valuable discussions with S. Bashinsky, E. Bertschinger, R. 
Bond, W. Hu, A. R. 
Liddle, D. H. Lyth, H. Martel, and P. Shapiro, and for help with integrals by M. Trott.  This article is based on work supported by the 
National Science Foundation under Grant No. 0071512, and also 
supported by The Robert A. Welch Foundation.

\clearpage

\begin{deluxetable}{cllll}
\tablecaption{Values of the transfer function for $\Omega_Mh^2=0.2$ and  
$\Omega_B/\Omega_M=0.01$ or $0.1$.}
\tablehead{
\colhead{$k\,({\rm Mpc}^{-1})$} &
\colhead{$T(k)_{\Omega_B/\Omega_M=0.01}$\tablenotemark{a}} & 
\colhead{$T(k)_{\Omega_B/\Omega_M=0.01}$\tablenotemark{b}} &
\colhead{$T(k)_{\Omega_B/\Omega_M=0.1}$\tablenotemark{a}} &
\colhead{$T(k)_{\Omega_B/\Omega_M=0.1}$\tablenotemark{b}}}
\startdata
0.1 & 0.161 & 0.0451 & 0.138 & 0.0509 \\
0.3 & 0.0398 & 0.0337 & 0.0328 & 0.0284 \\
0.5 & 0.0189 & 0.0169 & 0.0154 & 0.0140 \\
 1 & 0.00640   & 0.00586 & 0.00517 & 0.00480\\
2 & 0.00202 & 0.00187 & 0.00162 & 0.00152 \\
3 & 0.000997 & 0.000938 & 0.000797 & 0.000762
\enddata
\tablenotetext{a}{From Holtzman (1989)}
\tablenotetext{b}{From Eq.~(39)}

\end{deluxetable}

\end{document}